\documentclass[epj]{svjour}
\usepackage{graphics}
\usepackage{amsmath,amssymb}
\begin{document}
\title{Significance of self magnetic field in long-distance collimation of laser-generated electron beams}
\author{Shi Chen\inst{1} \and Jiaofeng Huang\inst{1} \and Yifei Niu\inst{1} \and Jiakun Dan\inst{1} \and Ziyu Chen\inst{1} \and Jianfeng Li\inst{1}
}                     
%
%
\institute{Institute of Fluid Physics, China Academy of Engineering Physics, Mianyang, Sichuan, 621900 China.}
\date{Received: date / Revised version: date}
%
\abstract{
Long-distance collimation of fast electron beams generated by laser-metallic-wire targets has been observed in recent experiments, while the mechanism behind this phenomenon remains unclear. In this work, we investigate in detail the laser-wire interaction processes with a simplified model and Classical Trajectory Monte Carlo simulations, and demonstrate the significance of the self magnetic fields of the beams in the long-distance collimation. Good agreements of simulated image plate patterns with various experiments and detailed analysis of electron trajectories show that the self magnetic fields provide restoring force that is critical for the beam collimation. By studying the wire-length dependence of beam divergence in certain experiments, we clarify that the role of the metallic wire is to balance the space-charge effect and thus maintain the collimation.
\PACS{
      {52.38.Kd}{Laser-plasma acceleration of electrons and ions}
     } 
} 
\maketitle
\section{Introduction}
\label{intro}
Energetic electron beams generated by ultra-intense laser pulses incident on solid-density targets have inspired wide interests in recent high-energy-density physics, due to their potential applications in table-top accelerators \cite{RevModPhys.81.1229}, advanced diagnostics \cite{RevModPhys.81.445} and fast ignition scheme for inertial confinement fusion \cite{PhysPlasmas.1.1626,Nature.412.798}. In order to fulfill these applications, it is critical to not only generate but also control such beams appropriately, such as collimating and guiding them over certain distances. Extensive researches have been carried out to generate collimated fast electron beams in laser-plasma interaction experiments. Collimated electron emission has been observed from/at the surface of a planar target over sub-millimeter distances \cite{PhysRevE.56.7179,PhysRevE.64.046407,PhysRevLett.81.112,PhysRevLett.84.674,PhysRevLett.96.165003,PhysRevLett.110.025001}. Hollow-cone and fibre-like devices \cite{Nature.432.1005,PhysPlasmas.16.020701,PhysRevE.86.065402,PhysRevLett.108.115004} have been designed to guide and collimate electron beams over distances of a few centimeters. Recently, spectacular advances have been made in long-distance collimation and guidance of electron beams by using a metallic wire as a guiding device \cite{PhysRevLett.106.255001,PhysRevLett.110.155001}. A significant number of fast electrons can be transported along the metallic wire over distances of tens of centimeters and even up to one meter, without a change in beam size. Such a metallic-wire device could provide an efficient means of energy transportation from an intense broad-energy electron source, such as a relativistic laser plasma.

Despite the experimental development of collimated electron beam generation, the mechanism underlying such long-distance collimation is still unclear. In models based on transient electric fields around wire targets, an extra, long-lived electric field is artificially introduced in order to interpret the observed collimation \cite{PhysRevLett.106.255001,PhysRevLett.110.155001}, but its origin is not explained clearly. On the other hand, the role of magnetic fields under such circumstances has long been ignored. However, it is well known that self magnetic fields could provide an important mechanism for fast-electron-beam collimation in other contexts. Both experimental \cite{PhysRevLett.110.025001,PhysRevE.64.046407,PhysRevLett.81.112,PhysRevLett.84.674} and theoretical \cite{PhysRevLett.82.743,PhysPlasmas.6.2855} researches have demonstrated the essential role of self magnetic fields on electron collimation in laser-solid-target interaction over short distances (e.g. tens of micrometers to several centimeters). In the context of dense magnetized plasmas \cite{PhysRevLett.103.105003,PhysPlasmas.12.032103} and astrophysical jets \cite{Nature.440.58,PhysRevLett.95.045002,PhysPlasmas.16.041005,PhysPlasmas.12.058301}, the collimating effect of self magnetic fields has also been verified over much longer distances (e.g. from tens of centimeters to several kilometers). Therefore, it is a natural question that how important the self magnetic fields could be in the long-distance electron collimation during laser-wire interaction processes. In this letter, we will investigate the significance of self magnetic field generated by electron beams and provide a new explanation to the mechanism of long-distance beam collimation in laser-metallic-wire interactions.
\section{Theoretical model}
\label{sec:1}
In the experiments where a metallic wire target being irradiated by an intense laser pulse, fast electron emission has been observed, as well as transient electric and magnetic fields. Because of electron emission, the metallic wire is positively charged and generates a strong, transient radial electric field. According to previous studies \cite{PhysRevLett.102.194801}, this electric field can be approximated by the following form:
\begin{equation}
    \label{equ:E}
    \mathbf{E}(r,t) = E_0\frac{r_0}{r}\exp{(-t/\tau_E)}\mathbf{e}_r, (0<z<L)
\end{equation}
where cylindrical coordinates $(r,\theta,z)$ are adopted and $\mathbf{e}_r$ is the unit vector in the radial direction. $E_0$ is the magnitude of the electric field on the wire surface, $r_0$ the radius of the wire, and $\tau_E$ is the relaxation time of the field. The field is assumed to be of importance only in the vicinity of the wire, i.e. $0<z<L$ where $L$ is the wire length. Outside this range the field is assumed to be zero. Under the effect of this attractive electric field, some of the emitted electrons either fall back into the wire or move helically along it, leading to a decrease in the electron beam divergence. However, such decrease in beam divergence due to only the electric field is not large enough to interpret the observed collimation, and that is why an extra long-lived electric field is artificially introduced in previous work \cite{PhysRevLett.106.255001}.

The magnetic fields generated in the process of laser-metallic-wire interaction mainly have two origins, i.e. the magnetic field generated by the return current and the self magnetic field of the electron beam. The existence of the return current has been demonstrated experimentally and theoretically \cite{Nature.432.1005,PhysPlasmas.11.2806,PhysPlasmas.17.056702,PhysPlasmas.19.113110,PhysRevLett.102.194801}. The magnetic field generated by this return current tends to divert electrons away from target surface, but generally can be balanced by the electric field. Moreover, the duration of return current so far has been proved to be no more than a few picoseconds, before it is dissipated in various ways. During such short time the fast electrons can only transport over a distance of less than one centimeter. Therefore, in the most time of long-distance transport, the effect of magnetic field due to return current can be neglected.

On the other hand, the self magnetic field of the electron beam could have significant influence on the motion of fast electrons, due to its much longer duration. It is well-known that fast-moving charged particle beams can generate considerable toroidal magnetic fields around them. The magnitude of this magnetic field can simply be estimated by Biot-Savart law. Assuming in a period of time $\tau_I$ there are $N$ electrons passing through, the corresponding current intensity $I = Ne/\tau_I$, where $e=1.6\times10^{-19}$ C is the unit charge. The magnetic field strength $B(r)=\mu_0I/2\pi r=\mu_0Ne/2\pi\tau_Ir$, where $\mu_0$ is the vacuum permeability and $r$ is the radial distance between the observation point and the current. Similar to the electric field, this magnetic field is also transient. Quinn {\it et al.} have studied these transient fields in the process of laser-metallic-wire interaction and observed that the typical variation time of the magnetic field is about tens of picoseconds \cite{PhysRevLett.102.194801}. Referring to their results, we assume that the magnetic field can be approximated by the following form:
\begin{equation}
    \label{equ:B}
    \mathbf{B}(r,t) = -B_0\frac{r_0}{r}\exp{(-t/\tau_B)}\mathbf{e}_\theta,
\end{equation}
where $\mathbf{e}_\theta$ is the unit vector along the angular direction. $B_0=\mu_0I/2\pi r_0$ is the magnetic field strength at the surface of the wire, and the minus sign indicates that the field points to the opposite direction of $\mathbf{e}_\theta$. $\tau_B$ is the relaxation time of the field. In the following, we will investigate the significance of such a magnetic field in long-distance collimation of fast electron beams.

\section{Numerical results}
\label{sec:2}
We carry out Classical Trajectory Monte Carlo (CTMC) simulations of electron trajectories under the influence of both the electric and self magnetic fields. The basic idea of our simulation is to solve the classical, relativistic equation of motion of fast electrons which has the following form:
\begin{equation}
    \label{equ:EOM}
    \frac{d\mathbf{p}}{dt} = -e(\mathbf{E}+\mathbf{v}\times\mathbf{B}),
\end{equation}
where $\mathbf{p}$ and $\mathbf{v}$ are electron momentum and velocity. The electric and self magnetic fields have the forms identical to Eq.(\ref{equ:E}) and (\ref{equ:B}), respectively. A fourth-order Runge-Kutta method is used to solve Eq.(\ref{equ:EOM}). The initial condition of each electron is randomly sampled by Monte Carlo method. All the electrons are assumed to be emitted near the laser spot, either from the fixed laser-incident point $(r_0,0,0)$ on the surface of the wire, or from the circumstance of the wire at $z=0$ and $r=r_0$. The initial kinetic energy and angular distributions are both uniform, in the range of $(E_{k,\text{min}},E_{k,\text{max}})$ and on a half-sphere, respectively. When the electrons hit the imaging plates (IPs) at a distance of $D$ from the electron source, their spatial positions in the $xy$-plane at $z=D$ are recorded to simulate the observed IP images in experiments.

\begin{figure}
    \resizebox{0.5\textwidth}{!}{%
  \includegraphics{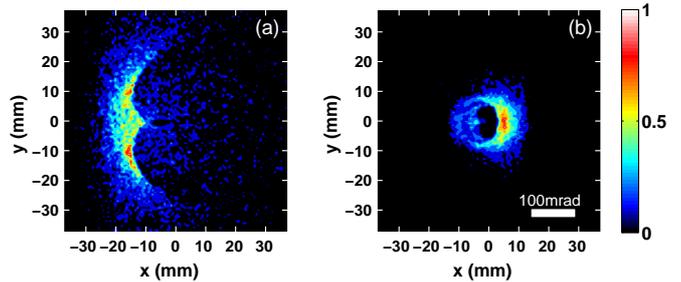}
}
    \caption{(color online) Electron-beam collimation with and without the self magnetic field. Simulated IP images with (a) only the electric field and (b) both the electric and self magnetic fields are numerically calculated. The magnitude of the electric field on the wire surface is $7\times10^8$ V/m, with a relaxation time of 5 ps. The self magnetic field strength at the edge of the beam is 0.5 T with a relaxation time of 50 ps. All the other parameters are identical.}
    \label{fig:1Evs1E1B}
\end{figure}

First of all, we study the importance of self magnetic field on electron beam collimation over a distance of tens of centimeters.We take the experiments carried out by Tokita, {\it et al.}  \cite{PhysRevLett.106.255001} as examples. In those experiments, an intense femtosecond laser pulse with peak intensity over $10^{18}$ W/cm$^2$ was incident onto the surface of a metallic wire target lying along the $z$-axis with a radius of $r_0=150$ $\mu$m. The emitted fast electron beams were collimated, moving along the wire target and were detected at a distance of $D=150$ mm by IP stacks. The detailed experimental setup and results can be referred to in Ref.\cite{PhysRevLett.106.255001}. In the case under discussion, the magnitude of the transient electric field on the wire surface is $E_0=7\times10^8$ V/m, with a relaxation time of $\tau_E=5$ ps, and the total number of emitted electrons $N$ is approximately $3\times10^9$ \cite{PhysRevLett.106.255001}. Since the duration of the laser pulse $\tau_L=150$ fs and the relaxation time of the electric field $\tau_E=5$ ps \cite{PhysRevLett.106.255001}, $\tau_I$ must be of a value between them, i.e. $\tau_L<\tau_I<\tau_E$ \cite{PhysRevLett.109.185001}. Therefore the typical magnetic field strength at the edge of the beam ($r\sim1$ mm) is roughly $0.02-0.7$ T. In our calculations it is assumed that $B_0=0.5$ T and $\tau_B=50$ ps. The initial position of each electron is fixed at the laser spot, i.e. $(r_0, 0, 0)$ in cylindrical coordinates. The initial range of electron kinetic energy is $50-1500$ keV. The wire length $L=30$ mm, and the electron positions are recorded at $D=150$ mm where they hit the IP stacks. All these parameters are identical to those in the experiments \cite{PhysRevLett.106.255001}. In order to investigate how important the self magnetic field could be, we compare the IP images of the electron beam with and without $\mathbf{B}$ in Fig.\ref{fig:1Evs1E1B}. When the self magnetic field is ignored, the beam is only weakly collimated with a vertical divergence of about 200 mrad (as shown in Fig.\ref{fig:1Evs1E1B}(a)). Non-negligible discrepancy between numerical and experimental results implies that a single electric field could not provide enough restoring force to collimate fast electrons. However, once taking into account of the self magnetic field, the divergence of the beam is reduced significantly to less than 70 mrad, as shown in Fig.\ref{fig:1Evs1E1B}(b). We also notice that the horizontal positions of the electron beams on the IP stacks are different, i.e. approximately $x=-18$ mm in Fig.\ref{fig:1Evs1E1B}(a) and $x=4$ mm in Fig.\ref{fig:1Evs1E1B}(b). Differences in both beam size and position show that the self magnetic field plays an important role in the process.

\begin{figure}
    \resizebox{0.5\textwidth}{!}{%
  \includegraphics{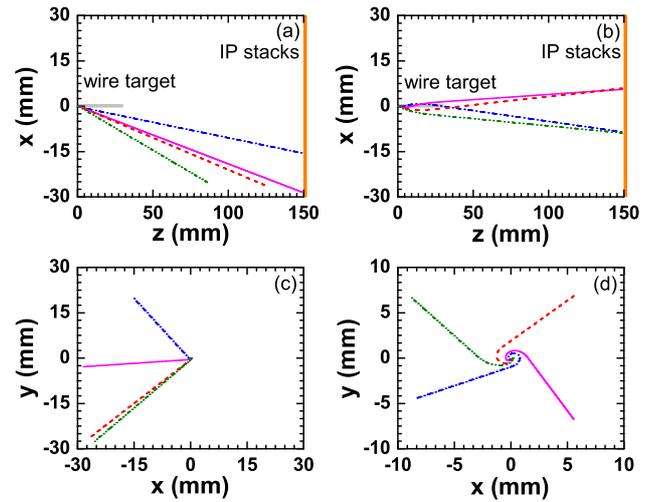}
}
    \caption{(color online) Comparison of electron trajectories with and without the self magnetic field. Four different line styles (i.e. solid, dash, dash-dotted and dash-dot-dotted curves) represent four different initial conditions of electrons. The short horizontal grey bar represents the wire target and the vertical orange bar indicates the IP stacks. The trajectories (a) with only the electric field and (b) with both electric and magnetic fields are projected onto the $xz$-plane. In the second row (i.e. (c) without and (d) with the magnetic field) these same trajectories are projected onto the $xy$-plane.}
    \label{fig:trace}
\end{figure}

To investigate in detail how electron motions are affected by the self magnetic field, we compare four typical electron trajectories with and without $\mathbf{B}$ in Fig.\ref{fig:trace}. The four different line styles represent four different initial conditions. The trajectories with only the electric field (see Fig.\ref{fig:trace}(a)) and with both electric and magnetic fields (see Fig.\ref{fig:trace}(b)) are projected onto the $xz$-plane. The short, horizontal grey bar is the wire target and the vertical orange bar represents the IP stacks. Without the self magnetic field (i.e. Fig.\ref{fig:trace}(a)), the distance between each electron and the wire target along the $x$-axis is greater than 15 mm at $z=150$ mm. Some of these electrons (e.g. the dashed and dash-dot-dotted lines) diverge so strongly that their $x$-direction distances from the wire become larger than 30 mm before they reach at the detector, which means that they cannot even be recorded by the IP stacks. However, under the influence of the self magnetic field (i.e. Fig.\ref{fig:trace}(b)), the divergence of these same electrons reduces significantly and all their $x$-direction distances are less than 10 mm from the wire, enabling them to hit around the central part of the IP stacks. In panel (c) and (d) of Fig.\ref{fig:trace} we project these trajectories onto the $xy$-plane. Panel (a) and (c) indicate that when ignoring the self magnetic field, the electron trajectories are nearly straight lines, while panel (b) and (d) show that after taking into account of the self magnetic field the trajectories become helical curves. This change in electron trajectories indicates that the magnetic field provides additional restoring force to these electrons and bends their trajectories more strongly towards the wire. By changing trajectories of individual electrons, the self magnetic field affects both the size and position of the beam, resulting its long-distance collimation.

\begin{figure}
    \resizebox{0.5\textwidth}{!}{%
  \includegraphics{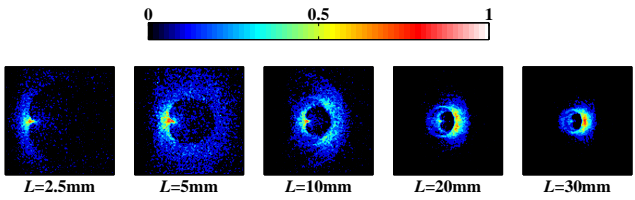}
}
    \caption{(color online) Wire-length dependence of the electron-beam collimation. The images of IP stacks for $L=2.5$, 5, 10, 20 and 30 mm are obtained by numerically calculating the electron trajectories under the influence of both electric and self magnetic fields. All the other parameters are kept identical.}
    \label{fig:1E1B5L}
\end{figure}

Another important feature of such a metallic-wire guiding device observed experimentally is the wire-length dependence of electron beam collimation \cite{PhysRevLett.106.255001}. As the wire length $L$ increases with other parameters kept unchanged, the beam divergence decreases significantly. In order to investigate this phenomenon, we carry out numerical calculations with different wire lengths, i.e. $L=2.5$, 5, 10, 20 and 30 mm, with other parameters unchanged and also taking into account of the self magnetic fields. The simulated IP images (as shown in Fig.\ref{fig:1E1B5L}) for all these wire lengths show clearly the ring-shaped pattern formed by the electrons. The size of the hole in the ring is nearly 60 mm in height for $L=2.5$ mm, and gradually decreases to less than 10 mm in height as $L$ increases to 30 mm. These results are in excellent agreement with those obtained in the experiments \cite{PhysRevLett.106.255001}. The mechanism behind this wire-length dependence of electron beam collimation is the balancing of space-charge effect by the metallic wire. After the beam leaves away from the wire but before it hits the IP stacks, the space-charge effect tends to diverge the electrons and destroy collimation. As the wire length increases, the duration of such space-charge effect decreases, leading to a higher degree of collimation. Therefore, the existence of the metallic wire balanced the space-charge effect and this is the reason for the observed wire-length dependence.

\begin{figure}
    \resizebox{0.5\textwidth}{!}{%
  \includegraphics{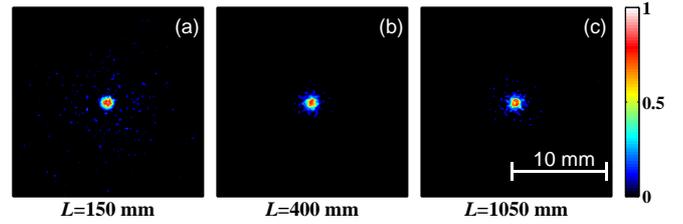}
}
    \caption{(color online) Long-distance collimation of the electron beams. The images of IP stacks are obtained by numerically calculating the electron trajectories under the influence of both electric and self magnetic fields for wire length of (a) 150 mm, (b) 400 mm and (c) 1050 mm. The typical magnitude of the electric field on the wire surface is $2.7\times10^{10}$ V/m with a relaxation time of 2 ps. The self magnetic field strength at the edge of the beam is 0.15 T. All the other parameters are kept identical.}
    \label{fig:long-dis}
\end{figure}

In order to further verify the role of self magnetic field in electron beam collimation, we investigate laser-wire interaction process over even longer distances, for example the experiments carried out by Nakajima, {\it et al.} \cite{PhysRevLett.110.155001}. In these experiments the wire target with a radius of $r_0=5$ $\mu$m is irradiated by an intense laser pulse with a peak intensity of about $10^{19}$ W/cm$^2$ and a duration of $\tau_L=150$ fs. The length of the wire varies from $L=150$ to 1050 mm. A stronger transient radial electric field is generated due to the stronger laser intensity as well as the smaller wire radius. The typical magnitude of this electric field on the wire surface $E_0$ is estimated to be about $2.7\times10^{10}$ V/m, with a relaxation time $\tau_E=2$ ps. The total beam charge is approximately 3 nC. The radius of the beam remains nearly unchanged for various wire lengths and is detected to be roughly 2 mm by IP stacks. Using the same simplified model mentioned above we estimate the typical self magnetic field strength at the edge of the beam is about $0.15-2$ T. In our simulations we set $B_0=0.15$ T and $\tau_B=50$ ps. The electrons are assumed to be emitted from the circumstance of the wire at $z=0$ and $r=r_0$. The initial angular distribution is uniform on a half-sphere, and the initial kinetic energy ranges from 50 to 1000 keV. The IP stacks are set at the end of the wire, recording electron positions in the $xy$-plane. In our calculations, three values of the wire length are used, i.e. $L=150$, 400 and 1050 mm. The simulated IP images for all the three cases are presented in Fig.\ref{fig:long-dis}. All the beams are in good collimation with a radius of roughly 2 mm, which agrees well with the experimental results. Different from the previous experiments, the beam radius remains nearly unchanged even when the length of the wire prolonged from 150 to 1050 mm, showing no obvious dependence of collimation on the wire length. Since in these experiments, the IP stacks are set at the end of the wire, the electrons never leave far from the vicinity of the wire. Therefore, the metallic wire keeps balancing the space-charge effect and helps to maintain beam collimation. In other words, these two types of experiments clarify that the self magnetic field of the beam provides the mechanism to collimate the electrons while the metallic wire provides a means to sustain the collimation.

\section{Conclusions}
\label{sec:3}
In summary, we investigate the long-distance collimation of fast electron beams in laser-metallic-wire interaction process through a simple model as well as numerical simulations, and demonstrate the significance of the self magnetic field of the beam in this long-distance collimation. Based on a simplified model, self magnetic fields with strength of roughly $0.1-1$ T are estimated to be generated according to experimental data. Numerical results of IP images of electron beams using Classical Trajectory Monte Carlo simulation agree well with various experimental configurations, indicating that the self magnetic field plays an important role in the divergence-free beam propagation. The relation between the beam collimation and the wire length is also analyzed, showing that the existence of the wire helps to maintain the beam collimation. Through our work the mechanism behind these long-distance beam collimation has been clarified.
%
%
%
%
 \bibliographystyle{unsrt}
 \bibliography{EPJD}

\providecommand{\noopsort}[1]{}\providecommand{\singleletter}[1]{#1}%
\begin{thebibliography}{10}

\bibitem{RevModPhys.81.1229}
E.~Esarey, C.~B. Schroeder, and W.~P. Leemans.
\newblock Physics of laser-driven plasma-based electron accelerators.
\newblock {\em Rev. Mod. Phys.}, 81:1229--1285, Aug 2009.

\bibitem{RevModPhys.81.445}
U.~Teubner and P.~Gibbon.
\newblock High-order harmonics from laser-irradiated plasma surfaces.
\newblock {\em Rev. Mod. Phys.}, 81:445--479, Apr 2009.

\bibitem{PhysPlasmas.1.1626}
Max Tabak, James Hammer, Michael~E. Glinsky, William~L. Kruer, Scott~C. Wilks,
  John Woodworth, E.~Michael Campbell, Michael~D. Perry, and Rodney~J. Mason.
\newblock Ignition and high gain with ultrapowerful lasers.
\newblock {\em Physics of Plasmas}, 1(5):1626--1634, 1994.

\bibitem{Nature.412.798}
R.~Kodama, P.~A. Norreys, K.~Mima, A.~E. Dangor, R.~G. Evans, H.~Fujita,
  Y.~Kitagawa, K.~Krushelnick, T.~Miyakoshi, N.~Miyanaga, T.~Norimatsu, S.~J.
  Rose, T.~Shozaki, K.~Shigemori, A.~Sunahara, M.~Tampo, K.~A. Tanaka,
  Y.~Toyama, T.~Yamanaka, and M.~Zepf.
\newblock Fast heating of ultrahigh-density plasma as a step towards laser
  fusion ignition.
\newblock {\em Nature}, 412(6849):798--802, August 2001.

\bibitem{PhysRevE.56.7179}
S.~Bastiani, A.~Rousse, J.~P. Geindre, P.~Audebert, C.~Quoix, G.~Hamoniaux,
  A.~Antonetti, and J.~C. Gauthier.
\newblock Experimental study of the interaction of subpicosecond laser pulses
  with solid targets of varying initial scale lengths.
\newblock {\em Phys. Rev. E}, 56:7179--7185, Dec 1997.

\bibitem{PhysRevE.64.046407}
Y.~T. Li, J.~Zhang, L.~M. Chen, Y.~F. Mu, T.~J. Liang, Z.~Y. Wei, Q.~L. Dong,
  Z.~L. Chen, H.~Teng, S.~T. Chun-Yu, W.~M. Jiang, Z.~J. Zheng, and X.~W. Tang.
\newblock Hot electrons in the interaction of femtosecond laser pulses with
  foil targets at a moderate laser intensity.
\newblock {\em Phys. Rev. E}, 64:046407, Sep 2001.

\bibitem{PhysRevLett.81.112}
M.~Borghesi, A.~J. MacKinnon, A.~R. Bell, R.~Gaillard, and O.~Willi.
\newblock Megagauss magnetic field generation and plasma jet formation on solid
  targets irradiated by an ultraintense picosecond laser pulse.
\newblock {\em Phys. Rev. Lett.}, 81:112--115, Jul 1998.

\bibitem{PhysRevLett.84.674}
R.~Kodama, K.~A. Tanaka, Y.~Sentoku, T.~Matsushita, K.~Takahashi, H.~Fujita,
  Y.~Kitagawa, Y.~Kato, T.~Yamanaka, and K.~Mima.
\newblock Long-scale jet formation with specularly reflected light in
  ultraintense laser-plasma interactions.
\newblock {\em Phys. Rev. Lett.}, 84:674--677, Jan 2000.

\bibitem{PhysRevLett.96.165003}
Y.~T. Li, X.~H. Yuan, M.~H. Xu, Z.~Y. Zheng, Z.~M. Sheng, M.~Chen, Y.~Y. Ma,
  W.~X. Liang, Q.~Z. Yu, Y.~Zhang, F.~Liu, Z.~H. Wang, Z.~Y. Wei, W.~Zhao,
  Z.~Jin, and J.~Zhang.
\newblock Observation of a fast electron beam emitted along the surface of a
  target irradiated by intense femtosecond laser pulses.
\newblock {\em Phys. Rev. Lett.}, 96:165003, Apr 2006.

\bibitem{PhysRevLett.110.025001}
S.~Chawla, M.~S. Wei, R.~Mishra, K.~U Akli, C.~D. Chen, H.~S. McLean,
  A.~Morace, P.~K. Patel, H.~Sawada, Y.~Sentoku, R.~B. Stephens, and F.~N. Beg.
\newblock Effect of target material on fast-electron transport and resistive
  collimation.
\newblock {\em Phys. Rev. Lett.}, 110:025001, Jan 2013.

\bibitem{Nature.432.1005}
R.~Kodama, Y.~Sentoku, Z.~L. Chen, G.~R. Kumar, S.~P. Hatchett, Y.~Toyama,
  T.~E. Cowan, R.~R Freeman, J.~Fuchs, Y.~Izawa, M.~H. Key, Y.~Kitagawa,
  K.~Kondo, T.~Matsuoka, H.~Nakamura, M.~Nakatsutsumi, P.~A. Norreys,
  T.~Norimatsu, R.~A. Snavely, R.~B. Stephens, M.~Tampo, K.~A. Tanaka, and
  T.~Yabuuchi.
\newblock Plasma devices to guide and collimate a high density of mev
  electrons.
\newblock {\em Nature}, 432(7020):1005--1008, December 2004.

\bibitem{PhysPlasmas.16.020701}
J.~A. King, K.~U. Akli, R.~R. Freeman, J.~Green, S.~P. Hatchett, D.~Hey,
  P.~Jamangi, M.~H. Key, J.~Koch, K.~L. Lancaster, T.~Ma, A.~J. MacKinnon,
  A.~MacPhee, P.~A. Norreys, P.~K. Patel, T.~Phillips, R.~B. Stephens,
  W.~Theobald, R.~P.~J. Town, L.~Van Woerkom, B.~Zhang, and F.~N. Beg.
\newblock Studies on the transport of high intensity laser-generated hot
  electrons in cone coupled wire targets.
\newblock {\em Physics of Plasmas}, 16(2):020701, 2009.

\bibitem{PhysRevE.86.065402}
K.~U. Akli, C.~Orban, D.~Schumacher, M.~Storm, M.~Fatenejad, D.~Lamb, and R.~R.
  Freeman.
\newblock Coupling of high-intensity laser light to fast electrons in
  cone-guided fast ignition.
\newblock {\em Phys. Rev. E}, 86:065402, Dec 2012.

\bibitem{PhysRevLett.108.115004}
T.~Ma, H.~Sawada, P.~K. Patel, C.~D. Chen, L.~Divol, D.~P. Higginson, A.~J.
  Kemp, M.~H. Key, D.~J. Larson, S.~Le~Pape, A.~Link, A.~G. MacPhee, H.~S.
  McLean, Y.~Ping, R.~B. Stephens, S.~C. Wilks, and F.~N. Beg.
\newblock Hot electron temperature and coupling efficiency scaling with
  prepulse for cone-guided fast ignition.
\newblock {\em Phys. Rev. Lett.}, 108:115004, Mar 2012.

\bibitem{PhysRevLett.106.255001}
Shigeki Tokita, Kazuto Otani, Toshihiko Nishoji, Shunsuke Inoue, Masaki
  Hashida, and Shuji Sakabe.
\newblock Collimated fast electron emission from long wires irradiated by
  intense femtosecond laser pulses.
\newblock {\em Phys. Rev. Lett.}, 106:255001, Jun 2011.

\bibitem{PhysRevLett.110.155001}
Hiroaki Nakajima, Shigeki Tokita, Shunsuke Inoue, Masaki Hashida, and Shuji
  Sakabe.
\newblock Divergence-free transport of laser-produced fast electrons along a
  meter-long wire target.
\newblock {\em Phys. Rev. Lett.}, 110:155001, Apr 2013.

\bibitem{PhysRevLett.82.743}
H.~Ruhl, Y.~Sentoku, K.~Mima, K.~A. Tanaka, and R.~Kodama.
\newblock Collimated electron jets by intense laser-beam\char21{}plasma surface
  interaction under oblique incidence.
\newblock {\em Phys. Rev. Lett.}, 82:743--746, Jan 1999.

\bibitem{PhysPlasmas.6.2855}
Y.~Sentoku, H.~Ruhl, K.~Mima, R.~Kodama, K.~A. Tanaka, and Y.~Kishimoto.
\newblock Plasma jet formation and magnetic-field generation in the intense
  laser plasma under oblique incidence.
\newblock {\em Physics of Plasmas}, 6(7):2855--2861, 1999.

\bibitem{PhysRevLett.103.105003}
Deepak Kumar and Paul~M. Bellan.
\newblock Nonequilibrium alfv\'enic plasma jets associated with spheromak
  formation.
\newblock {\em Phys. Rev. Lett.}, 103:105003, Sep 2009.

\bibitem{PhysPlasmas.12.032103}
S.~C. Hsu and P.~M. Bellan.
\newblock On the jets, kinks, and spheromaks formed by a planar magnetized
  coaxial gun.
\newblock {\em Physics of Plasmas}, 12(3):032103, 2005.

\bibitem{Nature.440.58}
Wouter H.~T. Vlemmings, Philip~J. Diamond, and Hiroshi Imai.
\newblock A magnetically collimated jet from an evolved star.
\newblock {\em Nature}, 440(7080):58--60, March 2006.

\bibitem{PhysRevLett.95.045002}
S.~You, G.~S. Yun, and P.~M. Bellan.
\newblock Dynamic and stagnating plasma flow leading to magnetic-flux-tube
  collimation.
\newblock {\em Phys. Rev. Lett.}, 95:045002, Jul 2005.

\bibitem{PhysPlasmas.16.041005}
P.~M. Bellan, M.~Livio, Y.~Kato, S.~V. Lebedev, T.~P. Ray, A.~Ferrari,
  P.~Hartigan, A.~Frank, J.~M. Foster, and P.~Nicola\"{i}.
\newblock Astrophysical jets: Observations, numerical simulations, and
  laboratory experiments.
\newblock {\em Physics of Plasmas}, 16(4):041005, 2009.

\bibitem{PhysPlasmas.12.058301}
P.~M. Bellan.
\newblock Miniconference on astrophysical jets.
\newblock {\em Physics of Plasmas}, 12(5):058301, 2005.

\bibitem{PhysRevLett.102.194801}
K.~Quinn, P.~A. Wilson, C.~A. Cecchetti, B.~Ramakrishna, L.~Romagnani,
  G.~Sarri, L.~Lancia, J.~Fuchs, A.~Pipahl, T.~Toncian, O.~Willi, R.~J. Clarke,
  D.~Neely, M.~Notley, P.~Gallegos, D.~C. Carroll, M.~N. Quinn, X.~H. Yuan,
  P.~McKenna, T.~V. Liseykina, A.~Macchi, and M.~Borghesi.
\newblock Laser-driven ultrafast field propagation on solid surfaces.
\newblock {\em Phys. Rev. Lett.}, 102:194801, May 2009.

\bibitem{PhysPlasmas.11.2806}
F.~N. Beg, M.~S. Wei, E.~L. Clark, A.~E. Dangor, R.~G. Evans, P.~Gibbon,
  A.~Gopal, K.~L. Lancaster, K.~W.~D. Ledingham, P.~McKenna, P.~A. Norreys,
  M.~Tatarakis, M.~Zepf, and K.~Krushelnick.
\newblock Return current and proton emission from short pulse laser
  interactions with wire targets.
\newblock {\em Physics of Plasmas}, 11(5):2806--2813, 2004.

\bibitem{PhysPlasmas.17.056702}
A.~J. Kemp, B.~I. Cohen, and L.~Divol.
\newblock Integrated kinetic simulation of laser-plasma interactions,
  fast-electron generation, and transport in fast ignition.
\newblock {\em Physics of Plasmas}, 17(5):056702, 2010.

\bibitem{PhysPlasmas.19.113110}
X.~H. Yang, M.~E. Dieckmann, G.~Sarri, and M.~Borghesi.
\newblock Simulation of relativistically colliding laser-generated electron
  flows.
\newblock {\em Physics of Plasmas}, 19(11):113110, 2012.

\bibitem{PhysRevLett.109.185001}
Shunsuke Inoue, Shigeki Tokita, Kazuto Otani, Masaki Hashida, Masayasu Hata,
  Hitoshi Sakagami, Toshihiro Taguchi, and Shuji Sakabe.
\newblock Autocorrelation measurement of fast electron pulses emitted through
  the interaction of femtosecond laser pulses with a solid target.
\newblock {\em Phys. Rev. Lett.}, 109:185001, Oct 2012.

\end{thebibliography}
%
%
%


\end{document}